\documentclass[12pt,twoside]{article}
\usepackage{graphicx}
\usepackage{amssymb}
\usepackage{amsmath}
\begin{document}

\title{Quantum Correlation Dynamics of Three Qubits in Non-Markovian Environments}

\author{ M. A.  Shukri$^{1,}$\footnote{E-mail: mshukri2006@gmail.com},  Fares S. S. Alzahri$^{1}$\footnote{E-mail: faresalzahri7@gmail.com}. and  Ali Saif M. Hassan$^{1,}$\footnote{E-mail: alisaif73@gmail.com}\\
$^1$ Department of Physics, University of Sana'a, Sana'a, Yemen.\\
$^2$ Department of Physics, University of Amran, Amran, Yemen.}
\maketitle

\begin{abstract}
We investigate the quantum correlation dynamics of three independent qubits each locally interacting with a zero temperature non-Markovian reservoir by using the Geometric measure of quantum discord (GQD). The dependence of quantum correlation dynamics on amount of non-Markovian, the degree of initial quantum correlation and purity of the initial states are studied in detail. It is found that the quantum correlation of such three qubits system revives after instantaneous disappearance period when a proper amount of non-Markovian is present. A comparison to the pairwise quantum discord and entanglement dynamics in three qubits system is also made.
\end{abstract}

\noindent{\it Keywords}: Geometric measure, Quantum discord, non-Markovian, Quantum Correlation Dynamics

\section{Introduction}

Understanding quantum correlations in a multipartite quantum state is a fundamental open problem. In quantum information theory, the problem of characterization of correlations present in a quantum state has been a fundamental problem generating intense research effort in the last two decades \cite{hor09,guh09}. Correlations in quantum states, with far-reaching implications for quantum information processing, are usually studied in the entanglement-versus-separability
scenario \cite{hor09} leading to important insights in quantum computing \cite{nie02}, quantum communication protocols like teleportation \cite{ben93,hor95}, superdense coding \cite{ben92}, cryptography \cite{gis02}, and so on. However, some results showed that quantum correlations cannot only be limited to entanglement, because separable quantum states can also have correlations which are responsible for the improvement of some quantum tasks that cannot be achieved by classical means \cite{kni98,dat07}. An alternative classification for correlations based on quantum measurements has arisen in recent years and also plays an important role in quantum information theory \cite{pia08,luo08}. This is the quantum-versus-classical paradigm for correlations. The first attempts in this direction were made by Ollivier and Zurek \cite{oll01} and by Henderson and Vedral \cite{hen01}, who studied quantum correlations from a measurement perspective and introduced quantum discord as a measure of quantum correlations which has generated increasing interest \cite{zur03,lan10}. Luo and Fu have suggested that the quantum discord $D(\rho)$ can be expressed alternatively as the minimal loss of correlations caused by the non-selective von Neumann projective measurement given by the set of orthogonal $1D$ projectors $\{\Pi_i^a\}$ acting on one part of the system \cite{luo10},
\begin{equation}\label{eq1}
	D(\rho)=\min_{\Pi^{a}}[I(\rho)-I(\Pi^{a}((\rho))],
\end{equation}
where
$$
\Pi^{a}(\rho)=\sum_{i}\left(\Pi_{i}^{a} \otimes I^{b}\right) \rho\left(\Pi_{i}^{a} \otimes I^{b}\right)
$$

Here the minimum is over the von Neumann measurements $\Pi^{a}=\left\{\Pi_{i}^{a}\right\}$ on a part say $a$ of a bipartite system $a b$ in a state $\rho$ with reduced density operators $\rho^{a}$ and $\rho^{b}$ and $\Pi^{a}(\rho)$ is the resulting state after the measurement. $I(\rho)=S\left(\rho^{a}\right)+S\left(\rho^{b}\right)-S(\rho)$ is the quantum mutual information, $ S(\rho)=-Tr(\rho \ln \rho)$ is the von Neumann entropy and $I^{b}$ is the identity operator on part $b$.\\
For tripartite and larger systems, several generalizations of discord have been proposed. In Ref. \cite{rul} a symmetric multipartite discord was deﬁned based on relative entropy and local measurements. Another deﬁnition of
multipartite discord was provided in Ref. \cite{okr}, as the sum of bipartite discords after making successive measurements. An approach using relative entropy was deﬁned in Ref. \cite{gio} to deﬁne genuine quantum and classical correlations in multipartite systems. Ref. \cite{cha} introduced the notion of quantum dissension deﬁned as the difference
between tripartite mutual information after a single measurement.\\
It is difficult to generalize the quantum discord in terms of quantum mutual information for multipartite cases \cite{cha,mod11}. To overcome this hurdle, Dakic \emph{et al} \cite{dak10}, have provided a geometric measure of quantum discord (GQD) as the distance for a given state to the closest classical quantum state
\begin{equation}\label{eq2}
	D(\rho)=\min_{\chi \in \Omega_0}||\rho-\chi||^2,
\end{equation}
with the minimum taken over the set $\Omega_{0}$ of zero discord states and where $\Omega_{0}$ is the set of classical-quantum states (set of zero-discord states $D(\chi)={0}$) and $\|\rho-\chi\|^{2}:=Tr(\rho-\chi)^{2}$ is the square Hilbert-Schmidt norm. Dakic \emph{et al} \cite{dak10} also obtained an explicit formula for GQD for a two qubit system based on the Hilbert-Schmidt distance. Consider a two-qubit state $\rho$ expressed in its Bloch representation as
$$
\rho=\frac{1}{4}\left(I^{a} \otimes I^{b}+\sum_{\alpha=1}^{3}\left(x_{\alpha} \sigma_{\alpha} \otimes I^{b}+I^{a} \otimes y_{\alpha} \sigma_{\alpha}\right)+\sum_{\alpha, \beta=1}^{3} t_{\alpha \beta} \sigma_{\alpha} \otimes \sigma_{\beta}\right)
$$
with $\{\sigma_{\alpha}\}$ being the Pauli operators. Then its geometric measure of quantum discord is given by \cite{dak10}
\begin{equation}\label{eq3}
	D(\rho)=\frac{1}{4}\left(\|\mathrm{x}\|^{2}+\|T\|^{2}-\lambda_{\max }\right)
\end{equation}
Here $\vec{x}:=\left(x_{1}, x_{2}, x_{3}\right)^{t}$ and $\vec{y}:=\left(y_{1}, y_{2}, y_{3}\right)^{t}$ are coherent (column) vectors for single qubit reduced density operators, $T=\left(t_{\alpha \beta}\right)$ is the correlation matrix, and $\lambda_{\max }$ is the largest eigenvalue of the matrix $\vec{x} \vec{x}^{t}+T T^{t} .$ The norms of vectors and matrices are the Euclidean norms, for example, $\|x\|^{2}:=\sum_{\alpha} x_{\alpha}^{2} .$ Here and throughout this paper, the superscript $t$ denotes the transpose of vectors and matrices and by the norm of any tensor, we mean its Euclidean
norm, that is, the square of the norm of a tensor is the sum of squares of its elements.\\
Hassan and Joag \cite{has12} provided an analytical formula for calculation of geometric measure of quantum discord for a $N$-qubit state corresponding to the von Neumann measurement on the $k$th part.

Recently, a natural generalization of quantum discord as originally defined in Ref. \cite{oll01,hen01} is made for multipartite systems using the concept of
conditional measurements, which  satisﬁes all of the postulates of a multipartite discord \cite{ben93}, but there is no analytical formula which can be used.

Generally, an actual quantum system is not closed and therefore will unavoidably be affected by surrounding environments \cite{bre02}. Due to the interaction with its environment, the quantum system is very fragile and easy to lose its quantum correlation, which is the main problem for the implementation of quantum information processing. In open composite quantum systems, the dynamic behavior of the correlations strongly depends on the noise generated by the surrounding environment. For a given quantitative system, the characterization of the environment as Markovian (no memory) or non-Markovian (with memory) is determined by the ratio between its typical correlation time and the system relaxation time. It is believed that the quantumness captured by discord is
different from entanglement \cite{oll01,zur03}. It was also investigated
under the Markovian environment in \cite{Wer09} for dissipative dynamics that the discord with an asymptotical
decrease is more robust than the entanglement with sudden death under the same conditions.
Wang at el \cite{wan10} studied the dynamics of the quantum discord by exactly solving a model consisting of two independent qubits subject to two zero-temperature non-Markovian reservoirs, respectively. This implies
that the quantum discord is more useful than the entanglement to describe the quantum correlation involved in
quantum systems. It was also investigated the dynamics of pairwise quantum discord (QD) for a mixed three-qubit $W$-type state in three independent non-Markovian reservoirs at zero temperature \cite{zha14}.

The paper is organized as follows. In sec. 2, we present the geometric measure of quantum correlation for multipartite systems. In sec. 3, we review the dynamics of a single qubit in non-Markovian environment, the procedure to solve the dynamics of three independent qubits is given and get the evolution of three-qubit density matrix. In sec. 4, we study the three-qubit quantum correlation dynamics in non-Markovian environment at zero temperature. A conclusion is given in sec. 5.

\section{Quantum Correlation}
Geometric measure has been used to measure the multipartite discord of quantum state with high dimensions.
Hassan and Joag \cite{has12} gave an exact computable formula for calculation of geometric measure of quantum discord for a $N$-qubit states. Following the same notation in Ref. \cite{has12},
\begin{equation}\label{eq4}
	\begin{array}{c}
		\sigma_{\alpha_{1}}^{(1)}=\left(\sigma_{\alpha_{1}} \otimes I_{2} \otimes \otimes I_{2}\right) \\
		\sigma_{\alpha_{2}}^{(2)}=\left(I_{2} \otimes \sigma_{\alpha_{2}} \otimes I_{2}\right) \\
		\sigma_{\alpha_{1}}^{(1)} \sigma_{\alpha_{2}}^{(2)}=\left(\sigma_{\alpha_{1}} \otimes \sigma_{\alpha_{2}} \otimes I_{2}\right)
	\end{array}
\end{equation}
We can write a three qubits state $\rho_{123}$ in the Bloch representation as

\begin{equation}\label{eq5}
\rho_{123}=\frac{1}{8}\{I_{8}+\sum_{k=1}^{3} \sum_{\alpha_{k}} s_{\alpha_{k}} \sigma_{\alpha_{k}}^{(k)}+\sum_{k \neq k^{\prime}=1}^{3} \sum_{\alpha_{k}, \alpha_{k^{\prime}}} t_{\alpha_{k} \alpha_{k^{\prime}}} \sigma_{\alpha_{k}}^{(k)} \sigma_{\alpha_{k^{\prime}}}^{(k^{\prime})}+\sum_{\alpha_{1} \alpha_{2} \alpha_{3}} t_{\alpha_{1} \alpha_{2} \alpha_{3}} \sigma_{\alpha_{1}}^{(1)} \sigma_{\alpha_{2}}^{\langle 2)} \sigma_{\alpha_{3}}^{(3)}\}
\end{equation}
where $\mathbf{s}^{(k)} (k=1,2,3) $ is a Bloch (coherent) vector corresponding to the $k$th qubit, $\mathbf{s}^{(k)}=\left[s_{\alpha_{k}}\right]_{\alpha_{k}=1}^{3}$, which is a tensor of order $1$ defined by
\begin{equation}\label{eq6}
	s_{\alpha_{k}}=Tr\left[\rho \sigma_{\alpha_{k}}^{(k)}\right]=Tr\left[\rho_{k}\sigma_{\alpha_{k}}\right],
\end{equation}
We denote the tensors of order $2$ by $T^{\{k, k'\}}=\left[t_{\alpha_{k}\alpha_{k'}}\right]$ which are defined by
\begin{equation}\label{eq7}
	t_{\alpha_{k} \alpha_{k^{\prime}}}=Tr\left[\rho_{123} \sigma_{\alpha_{k}}^{(k)} \sigma_{\alpha_{k^{\prime}}}^{\left(k^{\prime}\right)}\right]
\end{equation}
and the tensor of order $3$ by $\mathcal{T}=\left[t_{\alpha_{1}\alpha_{2} \alpha_{3}}\right],$ which are defined by
\begin{equation}\label{eq8}
	t_{\alpha_{1} \alpha_{2} \alpha_{3}}=Tr\left[\rho_{123} \sigma_{\alpha_{1}}^{(1)} \sigma_{\alpha_{2}}^{(2)} \sigma_{\alpha_{3}}^{(3)}\right]
\end{equation}
Then the geometric measure of quantum discord for a three qubits state corresponding to the von Neumann measurements $\{\widetilde{\Pi}^{(k)}\}$ on the $k$th qubit is given by,
\begin{equation}\label{eq9}
	D_{k}\left(\rho_{123}\right)=\frac{1}{8}\left\{\left\|\textbf{s}^{(k)}\right\|^{2}+\sum_{k^{\prime} \neq k=1}^{3}\left\|T^{\left\{k, k^{\prime}\right\}}\right\|^{2}+\|\mathcal{T}\|^{2}-\eta_{\max }^{(k)}\right\}
	; k=1,2,3.
\end{equation}
Here $\eta_{\max}^{(k)}$ is the largest eigenvalue of the matrix $G^{(k)}$ which is a $3 \times 3$ real symmetric matrix, defined as
\begin{equation}\label{eq10}
	G^{(k)}=\textbf{s}^{(k)}\left(\textbf{s}^{(k)}\right)^{t}+\sum_{k^{\prime} \neq k=1}^{3}\left(T^{\left\{k, k^{\prime}\right\}}\right)^{t}  T^{\left\{k, k^{\prime}\right\}}+\mathbf{T}^{(k)}
\end{equation}
where $\mathbf{T}^{(k)}=\left[\tau_{\alpha_{k} \beta_{k}}\right]$ is a $3 \times 3$ real matrix defined element wise as
$$\tau_{\alpha_{k} \beta_{k}}=\sum_{\alpha_{k_1}\alpha_{k_2}} t_{\alpha_{k_1}\alpha_{k_2} \alpha_{k} } t_{\alpha_{k_1}\alpha_{k_2} \beta_{k}}$$
$$\alpha_{k_1}, \alpha_{k_1},\alpha_{k} ,\beta_{k}=1,2,3;\;k_1,k_2=1,2,3\neq k.$$

Therefore, the geometric measure of quantum discord GQD (as represented bellow by $D_1,\; D_2,\; D_3$) of the successive measurement states and total quantum correlations TQC (as represented by $Q$) present in a three qubits state $\rho_{123}$ is given by \cite{has12},

$$D_1(\rho_{123})$$

$$D_2(\widetilde{\Pi}^{(1)}(\rho_{123}))$$

$$D_3(\widetilde{\Pi}^{(2)}(\widetilde{\Pi}^{(1)}(\rho_{123})))$$
\begin{equation}\label{eq11}
	Q(\rho_{123})= D_1(\rho_{123})+D_2( {\widetilde{\Pi}}^{(1)}(\rho_{123}))+D_3( {\widetilde{\Pi}}^{(2)}({\widetilde{\Pi}}^{(1)}(\rho_{123}))).
\end{equation}
The calculation of each term is shown in the appendix.
\section{MODEL}
We consider a system of three non-interacting qubits, that is, $A,\; B,$ and $C$, locally interacting with the reservoirs $R_A, \; R_B,$ and $R_C$, respectively, at  zero-temperature. The single qubit-reservoir pair is described by the following Hamiltonian
\begin{equation}\label{eq12}
	\hat{H}=\omega_{0} \sigma_{+} \sigma_{-}+\sum_{k}\left(\omega_{k} b_{k}^{\dag} b_{k}+g_{k} b_{k} \sigma_{+}+g_{k}^{*} b_{k}^{\dag} \sigma_{-}\right)
\end{equation}
with $\omega_{0}$ being the transition frequency of the qubit and $\sigma_{\pm}$ are the corresponding qubit's raining and lowering operators. The index $k$ labels different field modes of reservoir with frequencies $w_{k},$ $b_{k}^{\dag}$ and $b_{k}$ are the creation and annihilation operators with $g_{k}$ being the coupling constant to the qubit \cite{bre02,bel07,bel08}. This Hamiltonian represents one among the few open quantum systems amenable for an explicit solution \cite{gar97}. The dynamics of single qubit $s$ can be described by the reduced density matrix $\rho^{s}(t)$ which can be written, in the basis $\{|0\rangle,|1\rangle\}$ as \cite{bre02,bel07,man06}
\begin{equation}\label{eq13}
	\hat{\rho}^{s}(t)=
	\left(\begin{array}{cc}
		P_{\mathrm{t}} \rho_{11}^{s}(0) & \sqrt{P_{\mathrm{t}}} \rho_{\mathrm{10}}^{s}(0)\\
		\sqrt{P_{\mathrm{t}}} \rho_{01}^{s}(0) & \rho_{0 0}^{s}(0)+\left(1-P_{\mathrm{t}}\right) \rho_{11}^{s}(0)
	\end{array}\right)
\end{equation}
where the function $P_t$ obeys the differential equation
\begin{equation}\label{eq14}
	\dot{P}_{t}=-\int_{0}^{t} d t_{1} f\left(t-t_{1}\right) P_{t_{1}},
\end{equation}
and the correlation function is related to the spectral density $J(\omega)$ of the reservoir by $f\left(t-t_{1}\right)=\int d \omega J(\omega) e^{ i(\omega_{0}-\omega)(t-t_{1})}$. To get the exact form of $P_t$ thus depends on the choice of spectral density of the reservoir \cite{bre02}. In our model, we use the Lorentzian spectral distribution
\begin{equation}\label{eq15}
	J(\omega)=\frac{1}{2 \pi} \frac{\gamma_{0} \lambda^{2}}{\left(\omega_{0}-\omega\right)^{2}+\lambda^{2}}
\end{equation}
where $\lambda$ denotes spectral width of the coupling, it is related to the reservoir correlation time $\tau_{B}$ via $\tau_{B} \approx \lambda^{-1}$, while $\gamma_{0}$ is related to the decay of the atomic excited state in the Markovian limit of spectrum and is related to the qubit relaxation time by $\tau_{R} \approx \gamma_{0}^{-1}$. The relation between parameters $\gamma_{0}$ and $\lambda$ distinguishes between Markovian and non-Markovian regimes. In the Markovian regime there is $\gamma_{0}<\lambda / 2$ or $\tau_{R}>2 \tau_{B}$, and the non-Markovian regime corresponds to $\gamma_{0}>\lambda / 2$ or $\tau_{R}<2 \tau_{B},$ and the previously lost quantum information may be feedback into the system again. According to spectral density function $J(\omega)$ given in Eq.(\ref{eq15}), the solution of Eq.(\ref{eq14}) give
\begin{equation} \label{eq16}
P_{t}=e^{-\lambda t} [\cos(\frac{dt}{2}) +\frac{\lambda}{d}\sin(\frac{dt}{2})]^2,
\end{equation}
where $ d=\sqrt{2\gamma_{0}\lambda-\lambda^{2}}$ \cite{bre02,bel07}, which is an oscillating function that has discrete zeros $t=2[n\pi- \arctan(d/\lambda)]/d$ with $n$ being an arbitrary integer.

By using  the evolution of the reduced density matrix elements for the single qubit $\rho_{i\acute{i}}^s(t)=\sum_{l\acute{l}}A_{i\acute{i}}^{l\acute{l}}(t) \rho_{l\acute{l}}^s(0),$ to construct the reduced density matrix $\rho(t)$ for the three-qubit system as \cite{bel07,bel08},
\begin{equation}\label{eq17}
	\rho_{ii', jj',kk'}(t)= \sum_{ll',mm',nn'} A_{ii'}^{ll'}(t)B_{jj'}^{mm'}(t)C_{kk'}^{nn'}(t) \rho_{ll', m m', n n'}(0).
\end{equation}
By means of Eqs. (\ref{eq13}), (\ref{eq16}) and (\ref{eq17}) and under the standard product basis  $|1\rangle $=$ |111\rangle$,
$|2\rangle $=$ |110\rangle$, $ |3\rangle $=$ |101\rangle$,
$|4\rangle $=$ |100\rangle$, $ |5\rangle $=$ |011\rangle$,
$|6\rangle $=$ |010\rangle$, $ |7\rangle $=$ |001\rangle$,
$|8\rangle $=$ |000\rangle \rbrace$, we obtain the diagonal elements of the reduced density matrix for three-qubit
system as

\begin{eqnarray}\label{eq18}
	\rho_{11}(t) &=& P_{t}^{3} \rho_{11}(0), \nonumber\\
	\rho_{22}(t) &=& P_{t}^{2} \rho_{22}(0)+P_{t}^{2}\left(1-P_{t}\right) \rho_{11}(0), \nonumber\\
	\rho_{33}(t) &=& P_{t}^{2} \rho_{33}(0)+P_{t}^{2}\left(1-P_{t}\right) \rho_{11}(0), \nonumber\\
	\rho_{44}(t)&=& P_{t} \rho_{44}(0)+P_{t}\left(1-P_{t}\right) \rho_{33}(0)+P_{t}\left(1-P_{t}\right) \rho_{22}(0) 
	+ P_{t}\left(1-P_{t}\right)^{2} \rho_{11}(0), \nonumber\\
	\rho_{55}(t)&=& P_{t}^{2} \rho_{55}(0)+P_{t}^{2}\left(1-P_{t}\right) \rho_{11}(0), \nonumber\\
	\rho_{66}(t) &=& P_{t} \rho_{66}(0)+P_{t}\left(1-P_{t}\right) \rho_{55}(0)+P_{t}\left(1-P_{t}\right) \rho_{22}(0) +P_{t}\left(1-P_{t}\right)^{2} \rho_{11}(0), \nonumber\\
	\rho_{77}(t)&=& P_{t} \rho_{77}(0)+P_{t}\left(1-P_{t}\right) \rho_{55}(0)+P_{t}\left(1-P_{t}\right) \rho_{33}(0)+P_{t}\left(1-P_{t}\right)^{2} \rho_{11}(0), \nonumber\\
\rho_{88}(t)&=&\rho_{88}(0)+\left(1-P_{t}\right)\left[\rho_{77}(0)+\rho_{66}(0)+\rho_{44}(0)\right]
+\left(1-P_{t}\right)^{2}\left[\rho_{55}(0)+\rho_{33}(0)+\rho_{22}(0)\right]\nonumber\\
&& +\left(1-P_{t}\right)^{3} \rho_{11}(0) 
\end{eqnarray}
and the off-diagonal elements as:
\begin{eqnarray*}
	\rho_{12}(t) &=& P_{t}^{2} \sqrt{P_{t}} \rho_{12}(0) \nonumber\\
	\rho_{13}(t) &=& P_{t}^{2} \sqrt{P_{t}} \rho_{13}(0) \nonumber\\
	\rho_{14}(t) &=& P_{t}^{2} \rho_{14}(0) \nonumber\\
	\rho_{15}(t) &=& P_{t}^{2} \sqrt{P_{t}} \rho_{15}(0) \nonumber\\
	\rho_{16}(t) &=& P_{t}^{2} \rho_{16}(0) \nonumber\\
	\rho_{17}(t) &=& P_{t}^{2} \rho_{17}(0) \nonumber\\
	\rho_{18}(t) &=& P_{t} \sqrt{P_{t}} \rho_{18}(0) \nonumber\\
    \rho_{23}(t) &=& P_{t}^{2} \rho_{23}(0) \nonumber\\
	\rho_{24}(t) &=& P_{t} \sqrt{P_{t}} \rho_{24}(0)+P_{t} \sqrt{P_{t}}\left(1-P_{t}\right) \rho_{13}(0) \nonumber\\
	\rho_{25}(t) &=& P_{t}^{2} \rho_{25}(0) \nonumber\\
	\rho_{26}(t) &=& P_{t} \sqrt{P_{t}} \rho_{26}(0)+P_{t} \sqrt{P_{t}}\left(1-P_{t}\right) \rho_{15}(0) \nonumber\\
	\rho_{27}(t) &=& P_{t} \sqrt{P_{t}} \rho_{27}(0) \nonumber\\
	\rho_{28}(t) &=& P_{t} \rho_{28}(0)+P_{t}(1-P_t)\rho_{17}(0)\nonumber\\
	\end{eqnarray*}

\begin{eqnarray}\label{eq19}
	\rho_{34}(t) &=& P_{t}\sqrt{P_{t}} \rho_{34}(0)+P_{t}\sqrt{P_{t}}\left(1-P_{t}\right) \rho_{12}(0) \nonumber\\
	\rho_{35}(t) &=& P_{t}^{2} \rho_{35}(0) \nonumber\\
	\rho_{36}(t) &=& P_{t} \sqrt{P_{t}} \rho_{36}(0) \nonumber\\
	\rho_{37}(t) &=& P_{t} \sqrt{P_{t}} \rho_{37}(0)+P_{t} \sqrt{P_{t}}\left(1-P_{t}\right) \rho_{15}(0) \nonumber\\
	\rho_{38}(t) &=& P_{t} \rho_{38}(0)+P_{t}\left(1-P_{t}\right) \rho_{16}(0) \nonumber\\
	\rho_{45}(t) &=& P_{t} \sqrt{P_{t}} \rho_{45}(0) \nonumber\\
	\rho_{46}(t) &=& P_{t} \rho_{46}(0)+P_{t}\left(1-P_{t}\right) \rho_{35}(0) \nonumber\\
	\rho_{47}(t) &=& P_{t} \rho_{47}(0)+P_{t}\left(1-P_{t}\right) \rho_{25}(0) \nonumber\\
	\rho_{48}(t) &=& \sqrt{P_{t}} \rho_{48}(0)+\sqrt{P}_{t}\left(1-P_{t}\right) \rho_{37}(0) +\sqrt{P}_{t}\left(1-P_{t}\right) \rho_{26}(0)+\sqrt{P_{t}}(1-P_t)^2 \rho_{15}(0)\nonumber\\
	\rho_{56}(t) &=& P_{t} \sqrt{P_{t}}(1-P_t) \rho_{12}(0)+P_{t} \sqrt{P_{t}}\rho_{56}(0) \nonumber\\
	\rho_{57}(t) &=& P_{t} \sqrt{P_{t}}(1-P_t) \rho_{13}(0)+P_{t} \sqrt{P_{t}}\rho_{57}(0) \nonumber\\
	\rho_{58}(t) &=& P_{t} \rho_{58}(0)+P_{t} (1-P_{t})\rho_{14}(0) \nonumber\\
	\rho_{67}(t) &=& P_{t} \rho_{67}(0)+P_{t} (1-P_t)\rho_{23}(0) \nonumber\\
	\rho_{68}(t) &=& \sqrt{P_{t}}(1-P_t)^2 \rho_{13}(0)+\sqrt{P_{t}}(1-P_t)\rho_{24}(0)+\sqrt{P_{t}}(1-P_t) \rho_{57}(0)
		 + \sqrt{P_{t}}\rho_{68}(0) \nonumber\\
	\rho_{78}(t) &=& \sqrt{P_{t}}(1-P_t)^2 \rho_{12}(0)+\sqrt{P_{t}}(1-P_t)\rho_{34}(0)\nonumber\\
            & &+\sqrt{P_{t}}(1-P_t) \rho_{56}(0)+ \sqrt{P_{t}}\rho_{78}(0).
\end{eqnarray}

\section{Non-Markovian Quantum Correlation Dynamics}
For three-qubit states, there are two type of nonequivalent entangled states, which called $|GHZ\rangle$-state and $|W\rangle$-state \cite{hor09}, and any fully entangled three qubits state is  stochastic local operations and classical communication $SLOCC$-equivalent to either $|GHZ\rangle$ and $|W\rangle$ \cite{dur}. So, this states are important in quantum information and computation processing. We consider two initial states constructed with $|GHZ\rangle-$ and $|W\rangle-$like states,
\begin{equation} \label{eq20}
\rho^{\psi}=r |\psi\rangle\langle\psi|+\frac{1-r}{8} I_8, \; \rho^{\phi}=r |\phi\rangle\langle\phi|+\frac{1-r}{8} I_8.
\end{equation}
Here $r \; (0\leq r \leq 1)$ denotes the purity of the initial state and $|\psi\rangle=\alpha_{\psi} |000\rangle + \beta_{\psi} |111\rangle, \; |\phi\rangle=\alpha_{\phi} |001\rangle +\beta_{\phi} |010\rangle + \eta_{\phi} |100\rangle.$ It is obvious that the states $|\psi\rangle,\; |\phi\rangle$ are the $|GHZ\rangle$ and $|W\rangle$ respectively with $\alpha_{\psi(\phi)}$ real, $\beta_{\psi(\phi)}=|\beta_{\psi(\phi)}| e^{i\delta}$ and $\eta_{\phi}=|\eta| e^{i \epsilon}$ where $\alpha_{\psi}^2+|\beta_{\psi}|^2=1,\; \alpha_{\phi}^2+|\beta_{\phi}|^2+|\eta_{\phi}|^2=1.$ As is known the $|GHZ\rangle$ and $|W\rangle$ states are different classes of entanglement, we expect that the two initial states might have some difference in quantum correlation dynamics measured by geometric measures. Since, the exact expression of GQD and TQC are rather complicated and not very enlightening, we study only numerically of the quantum discord and total quantum correlation dynamics as follow.\\

\begin{figure}[h]
	\includegraphics[width=8cm,height=5cm]	{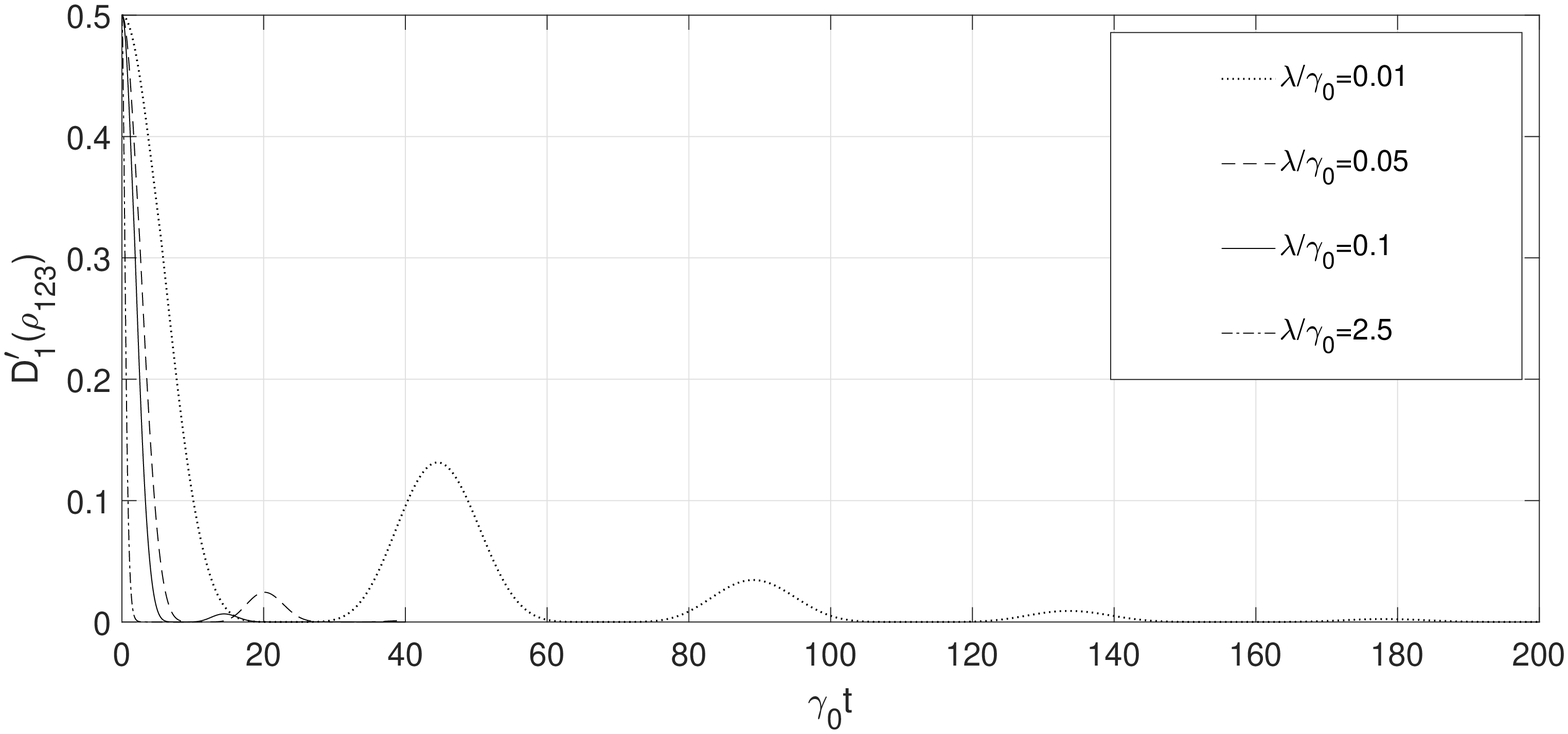}
	\includegraphics[width=8cm,height=5cm]	{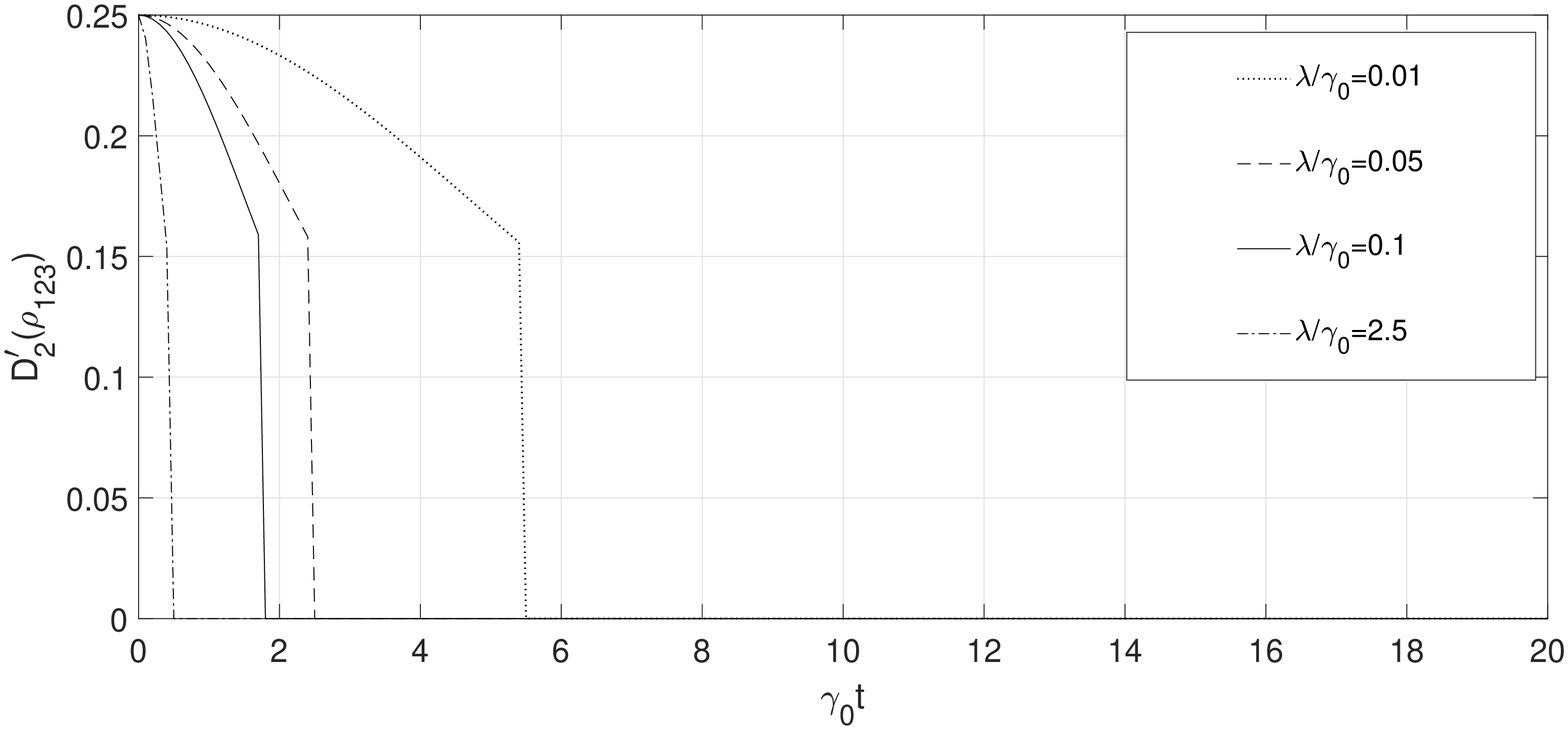}\\
	\includegraphics[width=8cm,height=5cm]	{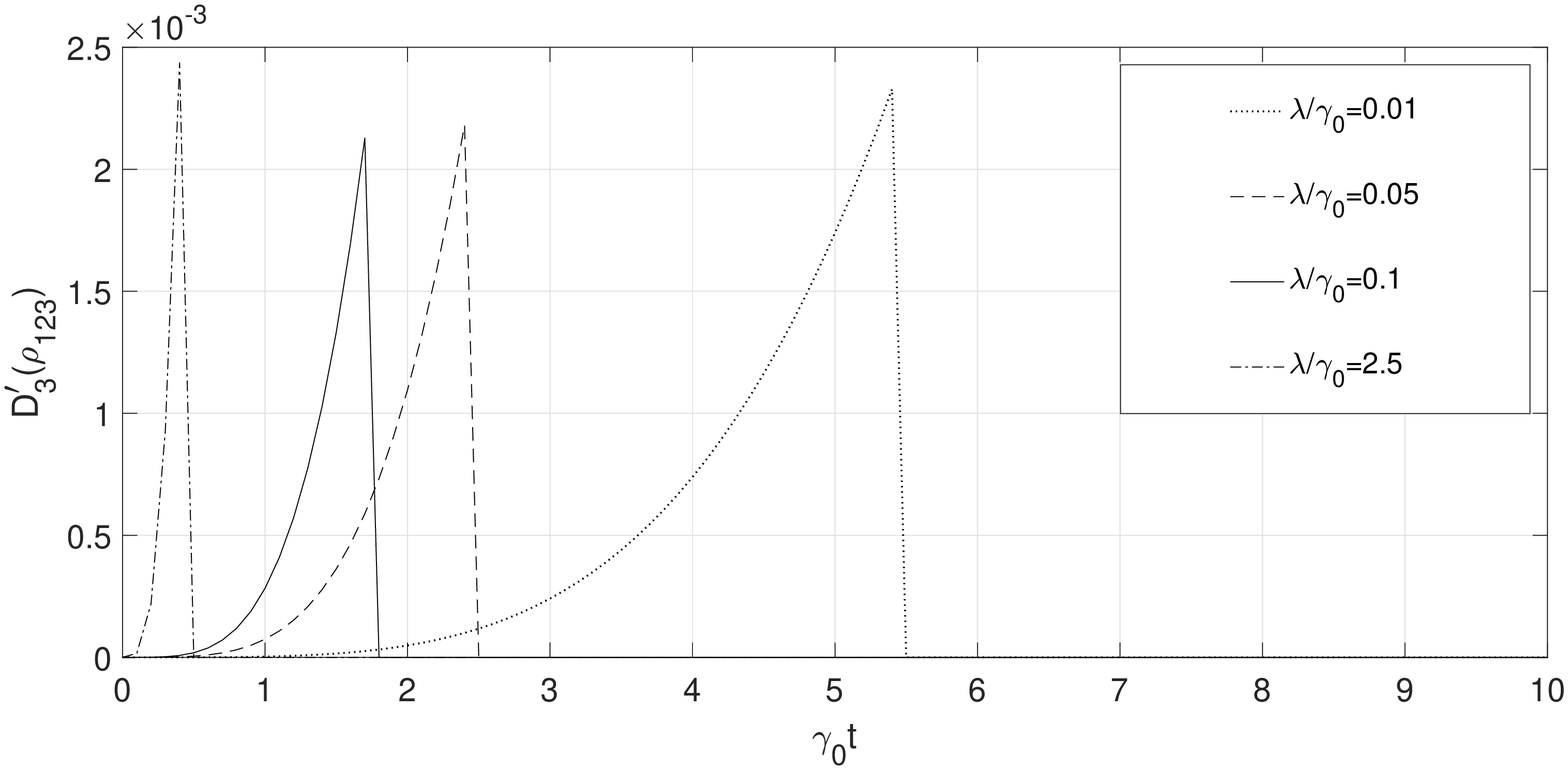}
	\includegraphics[width=8cm,height=5cm]	{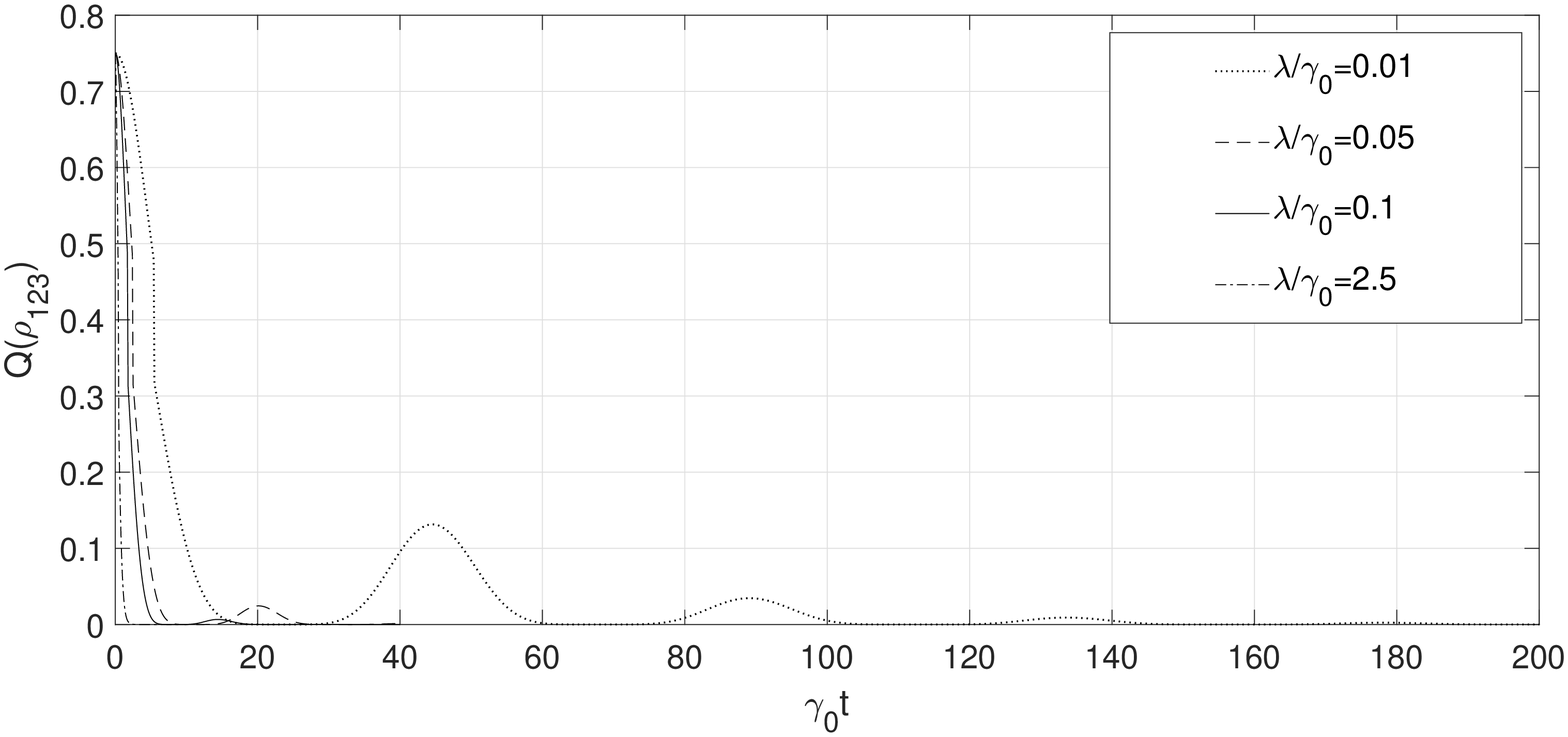}
	\caption{The geometric measures of QD $D_{1}\; \textbf{(a)}, \; D_{2}\; \textbf{(b)}, \; D_{3}\; \textbf{(c)},$ and $TQC \; \textbf{(d)}$ vs. $\gamma_0 t \; \textrm{for} \; \lambda / \gamma_{0}=2.5 \; \textrm{(dash-dotted line)},\; \lambda / \gamma_{0}=0.1 \; \textrm{(solid line)}, \;\lambda / \gamma_{0}=0.05 \;\textrm{(dashed line)},\; \lambda / \gamma_{0}=0.01 \; \textrm{(dotted line)}, \; \textrm{for} \; \rho^\psi \; \textrm{where} \; \alpha_{\psi}^2=\beta_{\psi}^2=1/2 \; \textrm{and} \; r=1.$}
	\label{fig:1}
\end{figure}

\begin{figure}[h]
	\includegraphics[width=8cm,height=5cm]	{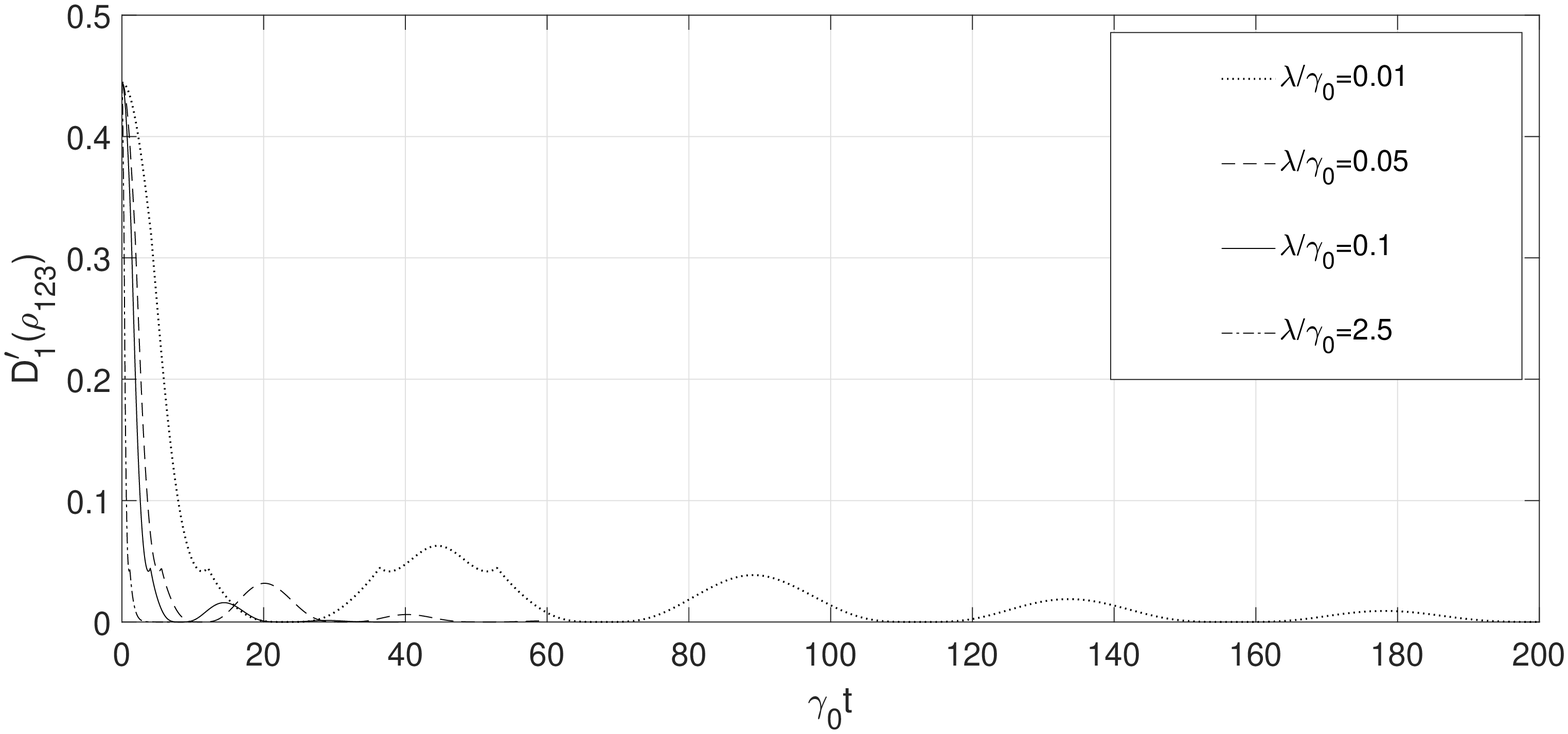}
	\includegraphics[width=8cm,height=5cm]	{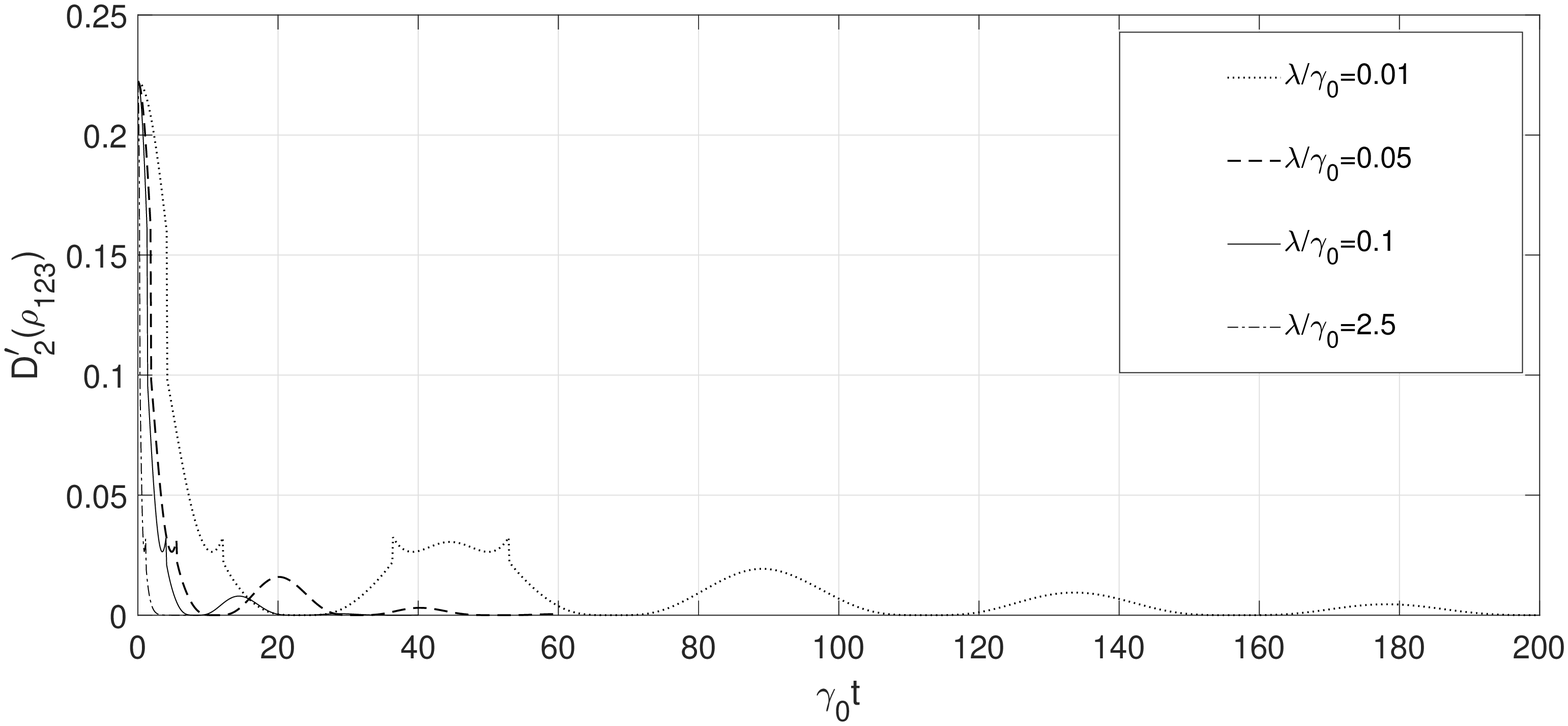}\\
	\includegraphics[width=8cm,height=5cm]	{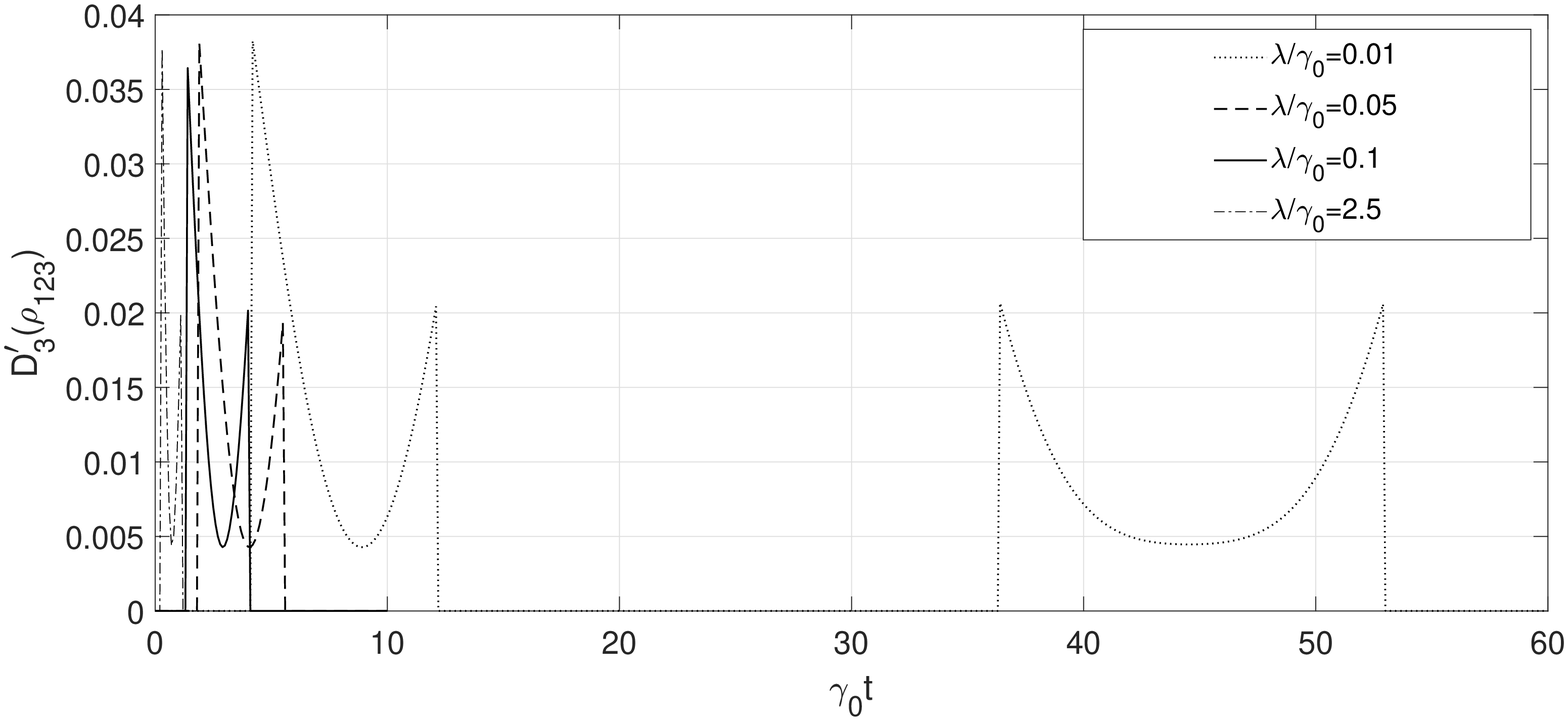}
	\includegraphics[width=8cm,height=5cm]	{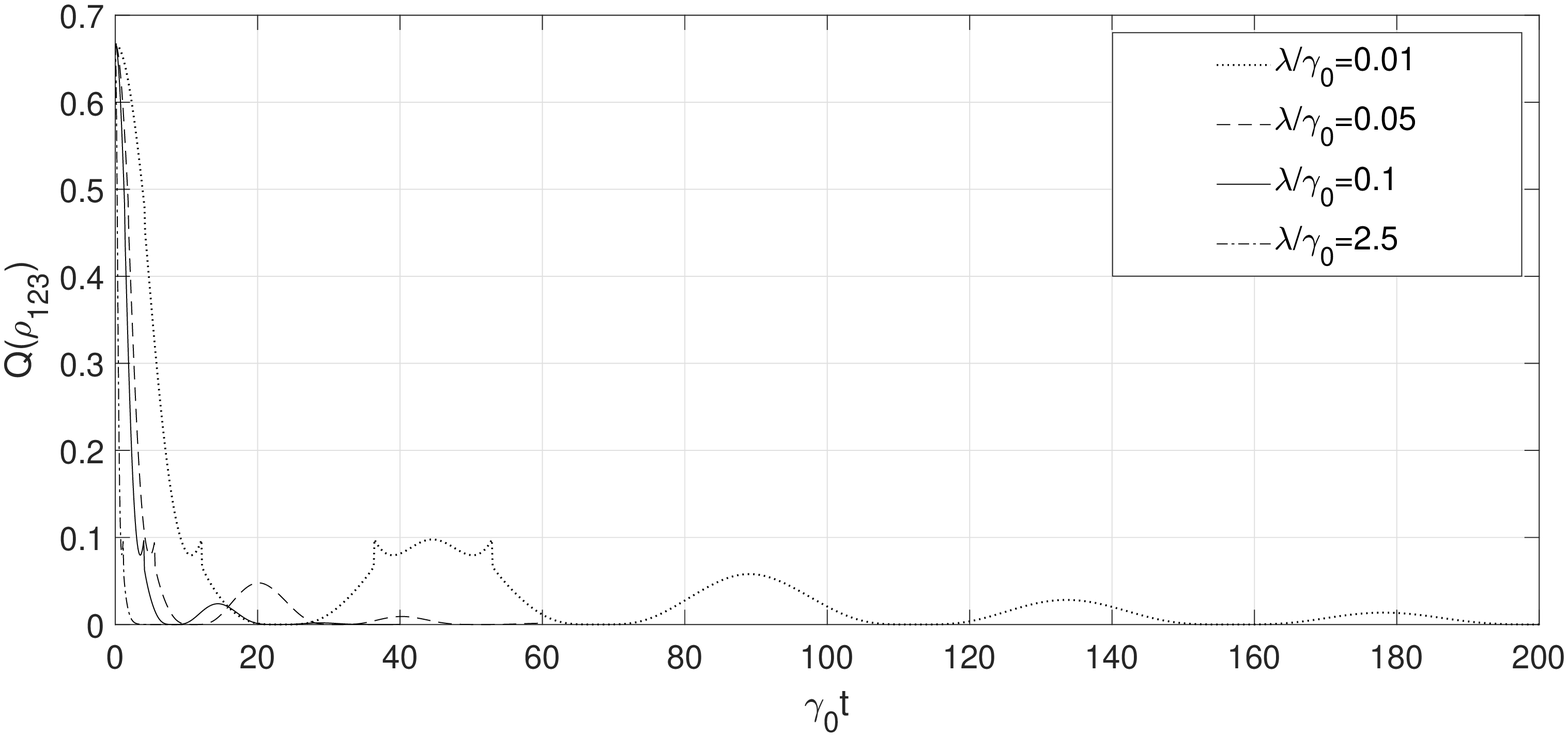}
	\caption{The geometric measures of QD $D_{1}\; \textbf{(a)},$ $D_{2}\; \textbf{(b)},$ $D_{3}\; \textbf{(c)},$ and $TQC \; \textbf{(d)}$  vs, $\gamma_{0} t$ for $\lambda / \gamma_{0}=2.5$ (dash-dotted line), $\lambda / \gamma_{0}=0.1$ (solid line), $\lambda / \gamma_{0}=0.05$ (dashed line), $\lambda / \gamma_{0}=0.01$ (dotted line), for $\rho^\phi$  where $\alpha_{\phi}^2= \beta_{\phi}^2=\eta_{\phi}^2=1/3, \; r=1 .$}
	\label{fig:2}
\end{figure}

\noindent\emph{\textbf{Case 1}} In this case we mainly study how the GQD and TQC dynamics is influenced by the degree of non-Markovian. To this purpose, we plot, Figs. \ref{fig:1}, \ref{fig:2}, the GQD and TQC as functions of $\gamma_0 t$ for four values of $\lambda/\gamma_0$ ($\lambda/\gamma_0 = 2.5, 0.1,0.05,0.01$), starting from pure $\rho^{\psi},\; \rho^{\phi}$ states, that is, for $r=1$ and taking $\alpha_{\psi}^2=\beta_{\psi}^2=1/2,\; \alpha_{\phi}^2= \beta_{\phi}^2=\eta_{\phi}^2=1/3.$
This choice of the parameters permits us to evidence quite different time behaviors of GQD and TQC in terms of $\lambda/\gamma_0$ for the two initial states $\rho^{\psi}(0)=|\psi\rangle\langle\psi|$ and $\rho^{\phi}(0)=|\phi\rangle\langle\phi|.$
Geometric measure of quantum discord and total quantum correlation (as stated in Eq. (\ref{eq11})) of state $|\psi\rangle$, under this choice of parameters, as a function of $\gamma_0 t$ for the four values of $\lambda/\gamma_0$ is plotted in Figs. \ref{fig:1}. The counter part of state $|\phi\rangle$ are correspondingly plotted in Figs. \ref{fig:2}. According to the conditions for Markovian and non-Markovian dynamics regimes, Markovian dynamics take place when $\lambda/\gamma_0 =2.5$, while $\lambda/\gamma_0 =0.1,\;0.05,\; 0.01$ the non-Markovian dynamics is relevant.

Figures \ref{fig:1} and \ref{fig:2} show three similar characters. Firstly, the geometric measures of quantum discord and total quantum correlation, measured by $D_{1}(\rho_{123})$ and $Q(\rho_{123})$ respectively, evolution in non-Markovian regime differs in essence from that in Markovian regime. Geometric measures in Markovian regime vanishes only in an asympotical way without revival. whereas, geometric measures in non-Markovian regime ($\lambda/\gamma_0 =0.1,\;0.05,\; 0.01$) decays gradually to zero but reappears after a period of time with a damping amplitude, which is in good agreement with the corresponding result described in Ref. \cite{zha14} for three qubits and in Ref. \cite{wan10} for two qubits. In contrast to the Figs. 1, 2 in Ref. \cite{zha} the entanglement in Markovian regime decays exponentially to zero and finally vanishes. However, entanglement in non-Markovian regime also decays to zero but reappears after a period of time with a damping amplitude. As there is no interaction between three qubits initially, this revival phenomenon is due to single qubit non-Markovian dynamics resulting from the memory effect of the environment. Secondly, the non-Markovian decay rate and revival amplitude depend on the amount of non-Markovian $\lambda/\gamma_0$ with  $\lambda/\gamma_0$ decreasing, which means degree of non-Markovian increasing, non-Markovian decay rate decreases and the revival amplitute increases and the numbers of revival amplitude increases with degree of non-Markovian increasing. Thirdly, there is no decay rate of $D_3$ of both Figs. 1c (for GHZ-state) and 2c (for  W-state) and it has anomalous behavior different of that for $D_{1}$ and $D_{2}$ evolution.\\


Moreover, Figures \ref{fig:1} and \ref{fig:2} show three different characters. Firstly,  the number of revival amplitude of Fig. \ref{fig:2} (for W-state) is more than that of Fig. \ref{fig:1} (for GHZ-state) in non-Markovian dynamics regime, but the maximum values of the revival amplitudes of  Fig. \ref{fig:2} (for W-state) is smaller than that of Fig. \ref{fig:1} (for GHZ-state). Secondly, the decay rate of Fig. 1b, $D_2$, (for GHZ-state), is faster than that of Fig. 2b (for W-state) and there is no revival amplitude of Fig. 1b, $D_2$, in non-Markovian dynamics regime for GHZ-state, while there is of Fig. 2b (for W-state). Thirdly, the anomalous behavior of Fig. 1c (for GHZ-state) differ from that of Fig. 2c (for W-state), the revival amplitude reappears once in Fig. 1c, while in Fig. 2c it reappears more than one and differ in their form.\\

\begin{figure}[h]
		\includegraphics[width=8cm, height=5cm]{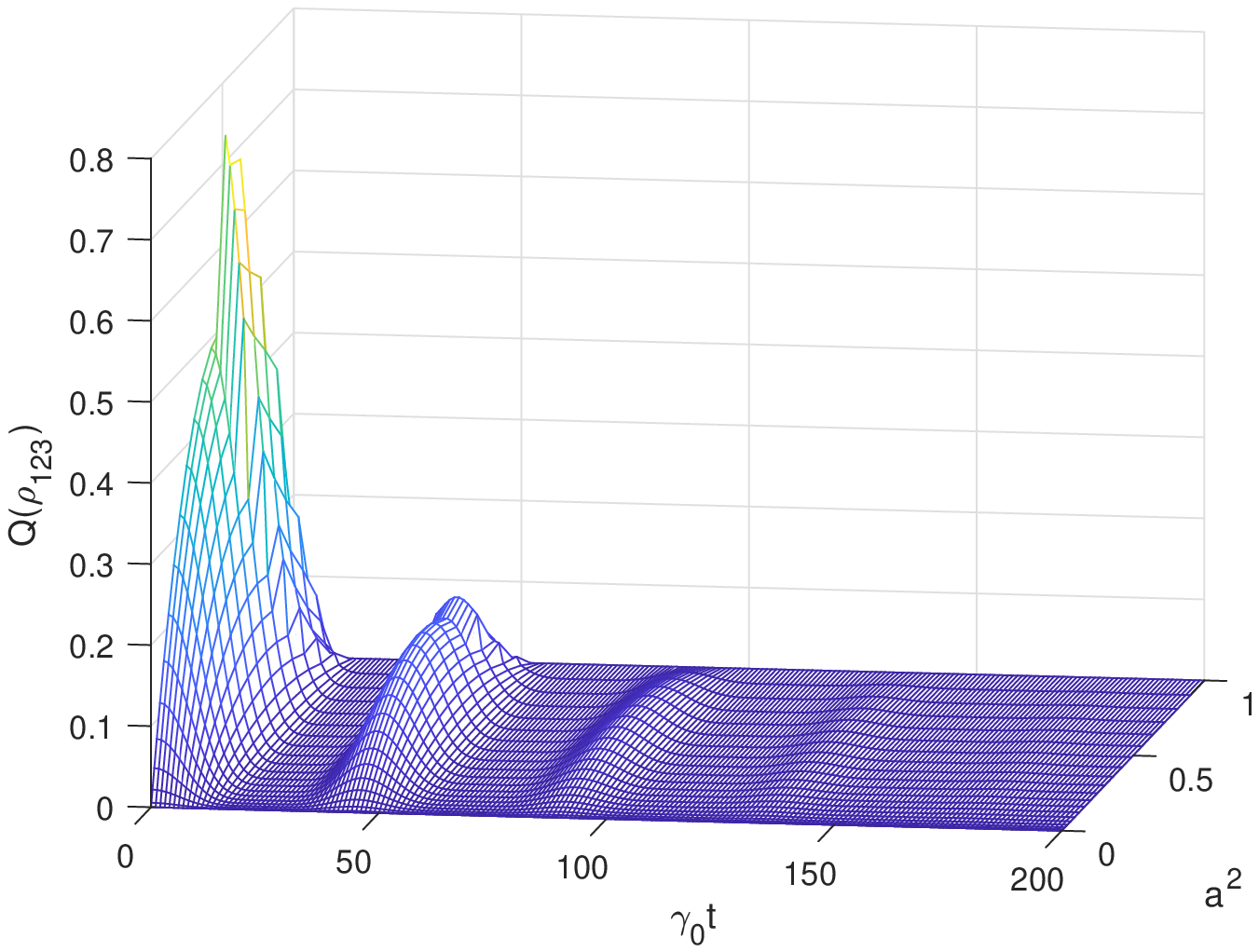}
		\includegraphics[width=8 cm, height=5cm]{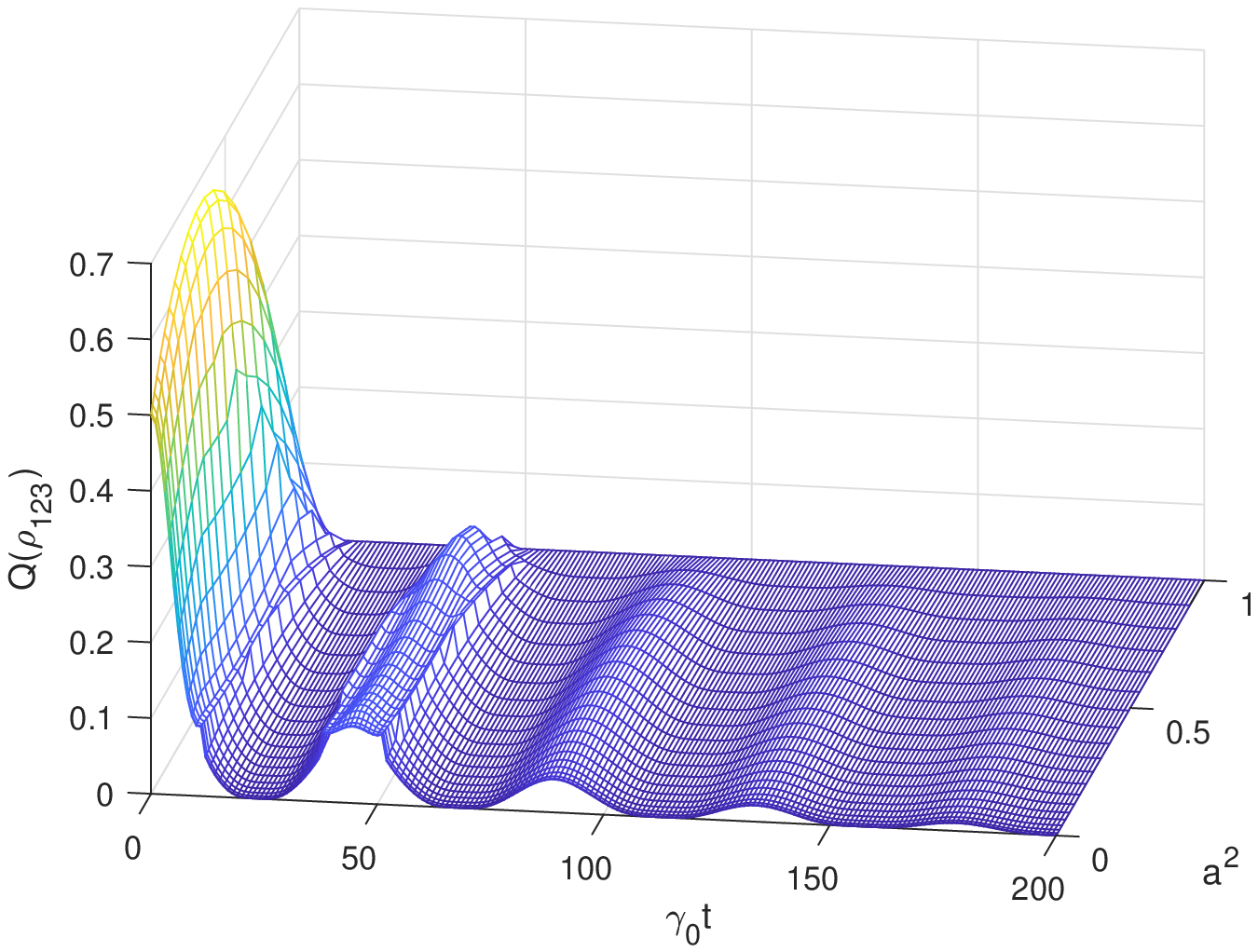}
		\caption{TQC, ($Q(\rho_{123})$) in terms of $\gamma_{0} t$ and $\alpha_{\psi(\phi)}^{2}$ for the initial GHZ-state left panel and W-state right panel, parameter $\lambda / \gamma_{0}=0.01,$  where $\beta_{\psi}=\sqrt{1-\alpha_{\psi}^{2}},$ $\beta_{\phi}=\eta_{\phi}=\sqrt{\left(1-\alpha_{\phi}^{2}\right) / 2}.$}
	\label{fig:3}
\end{figure}

\noindent\emph{\textbf{Case 2}} Another aspect of interest is how the three qubits geometric measure of TQC dynamics is affected by the degree of initial quantum correlation represent by $\alpha_{\psi(\phi)},\; \beta_{\psi(\phi)}$ and $\eta_{\phi}$. We assume the initial states with: $r=1, \; \beta_{\psi}=\sqrt{1-\alpha_{\psi}^2}, \; \beta_{\phi}=\eta_{\phi}=\sqrt{(1-\alpha_{\phi}^2)/2}$ and plot TQC as a function of $\gamma_0 t$ and $\alpha^2$ in Fig. \ref{fig:3}. Figure \ref{fig:3} presents the case for the non-Markovian regime $(\lambda/\gamma_0=0.01).$ We plot in Fig. 3-left panel and Fig. 3-right panel for states $\rho^{\psi}$ (GHZ-state) and $\rho^{\phi}$ (W-state) respectively. It can be seen that The TQC of both $\rho^{\psi}=|\psi\rangle\langle\psi|$ and $\rho^{\phi}=|\phi\rangle\langle\phi|$ periodically vanishes in accordance with the zero points of the function $P_t$ following the asymptotically damping. We can also see that the revival amplitudes of TQC increases with $\alpha^2$ increasing from $0$ to $1/2$, reaching its maximum value at $\alpha^2=1/2$, and decreases with $\alpha^2$ increasing from $1/2$ to $1$ in Fig. 3-left panel, while Fig. 3-right panel show that $TQC$ revival amplitude occurs in all the region of $\alpha^2$ with periodically amplitude damping. Comparing influence of initial quantum correlation on TQC dynamics in three qubits system to that in two qubits system  \cite{wan10}, we find that the effects are in general similar. Comparing entanglement dynamics in Figs. 3, 4 of Ref. \cite{zha14} to Figs. 3 in our work, we find that entanglement, for GHZ-state as measured by negativity, shows entanglement sudden death (ESD) occurs when $0\leq\alpha^2 \leq 1/2$ and the dark period decreases with $\alpha^2$ increasing, while in Fig. 3-left panel in our work the dark period does not occur in the whole region of $\alpha^2$. Besides, Fig. 4 of Ref. \cite{zha14} (for W-state) ESD does not occur in the all region of $\alpha^2$, whereas Fig. 3-right panel of this paper show similar behavior but revival amplitude is larger and zero period is smaller. \\

\noindent\emph{\textbf{Case 3}} In this case we analyze how the three qubits TQC dynamics is affected by the presence of mixedness in initial states regulated by the purity parameter $r$. We set the parameters in two initial states as $\alpha_{\psi}^2=\beta_{\psi}^2=1/2,\; \alpha_{\phi}^2=\beta_{\phi}^2=\eta_{\phi}^2=1/3$ and plot the geometric measure of TQC as a function of $\gamma_0 t$ and $r$ in Fig. 4-left panel and Fig. 4-right panel for GHZ and W states respectively.

Figures 4-left panel, 4-right panel show that TQC increases with purity parameter $r$ increasing at $t=0$ for both states. Also, it con be seen that the value and period of revival amplitude increase with purity $r$ increasing and when $r=1$ the revival amplitudes get the highest value with largest period.

Contrary, the zero periods of TQC increase with purity $r$ decreasing and when $r=0$ the zero periods are the largest. The difference between two figures is  the top of first revival amplitude in Fig.4-left panel is norm, while in Fig.4-right panel is warped and the warped revival amplitude increase with $\lambda/\gamma_0$ increasing.
Compared with Fig. 3 of Ref. \cite{zha14} of pairwise quantum discord dynamics for W-state, Fig. 4-right panel in this paper show similar behaviors. In contrast to Fig. 7 and Fig. 8 of Ref. \cite{zha} in non-Markovian regime, the entanglement dynamics sudden death occurs in both states in almost all the region of $r$, the TQC in Figs. 4-left panel and 4-right panel of our paper show that no dark period in nearly all region of $r$ for both same initial states. This phenomenon is further evidence that TQC is more robust than the entanglement against decoherence.
\begin{figure}[h]
	\includegraphics[width=8 cm, height=5cm]{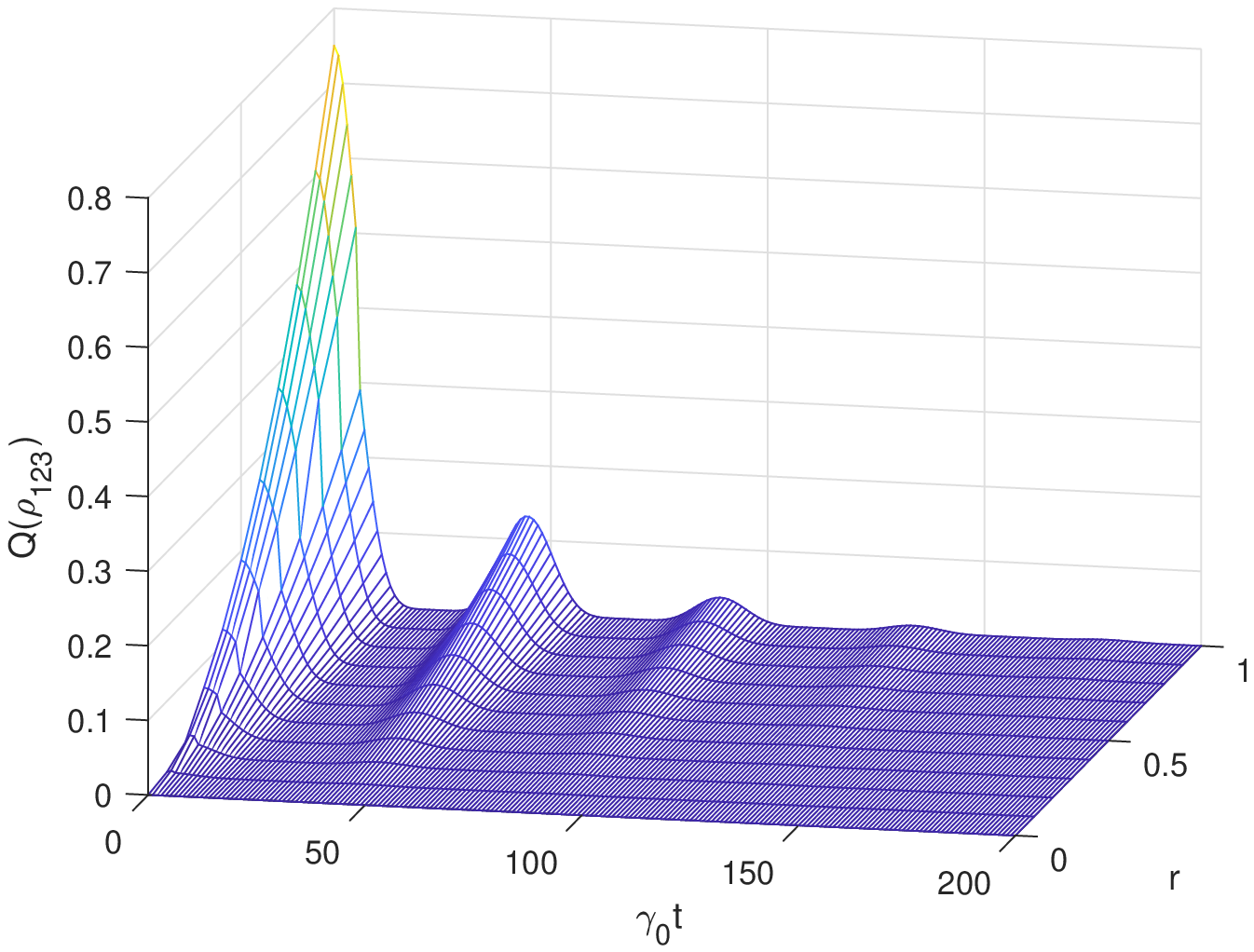}
	\includegraphics[width=8 cm, height=5cm]{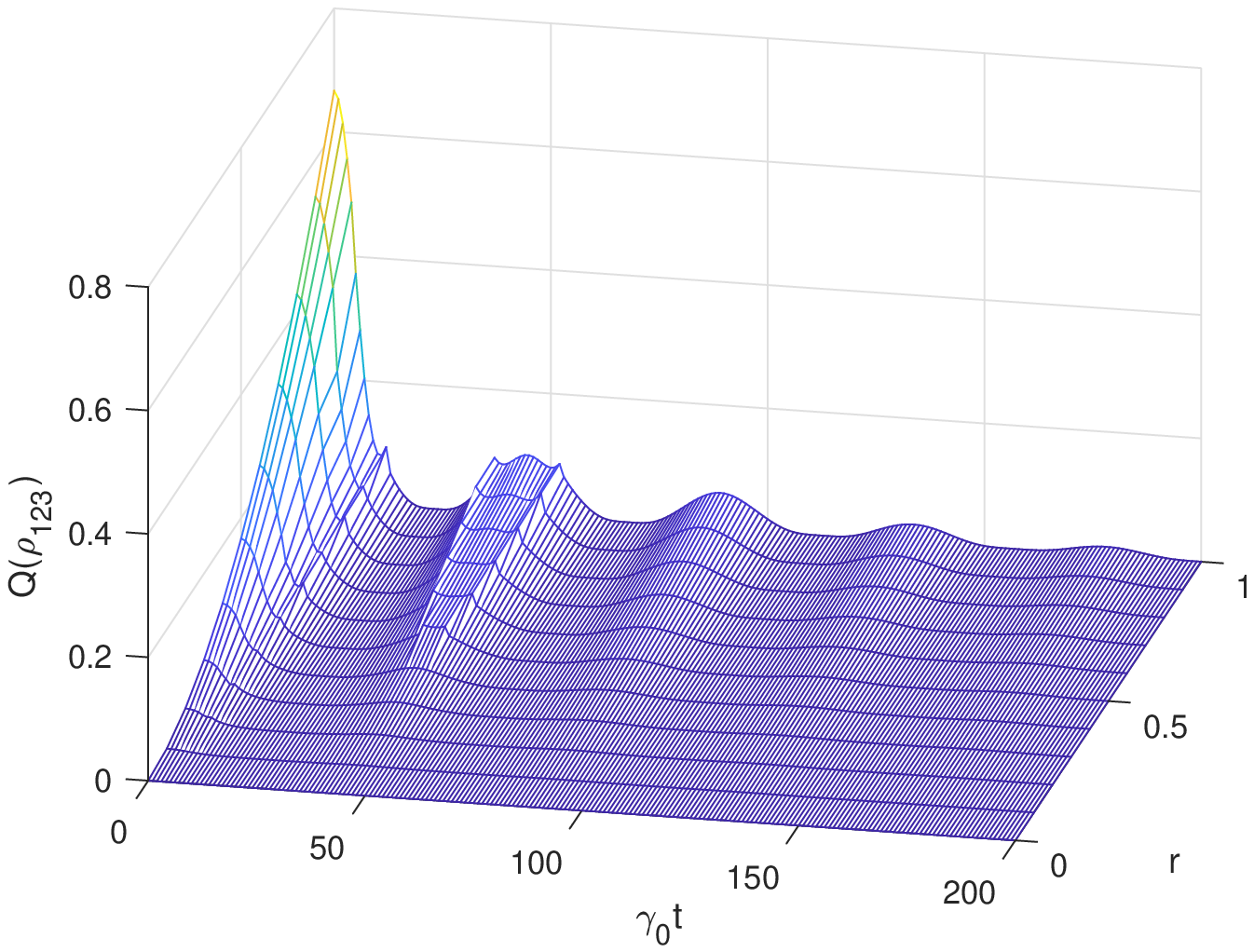}
	\caption{TQC, $Q(\rho_{123})$ in terms of $\gamma_{0} t$ and $r$ for the initial GHZ-state left panel and W-state right panel, parameter $\lambda / \gamma_{0}=0.01.$}
	\label{fig:4}
\end{figure}

\section{Conclusion}
We have studied the dynamics of quantum correlation of three qubits using exactly solvable model where each qubit independently and locally interacts with zero-temperature reservoir. We have discussed the different effects from the Markovian and non-Markovian reservoir. Specially, in the first case, we have analyzed the effects of the amount of  non-Markovian $\lambda/\gamma_0$ on the geometric measure of quantum discord in detail. We obtain that the two types initial states have some similar and different characters. The non-Markovian decay rate  decreases and revival amplitude, in both its value and numbers, increases with the amount of non-Markovian increasing. Moreover, there is no decay rate of $D_3$ for both initial GHZ and W states and $D_3$ has anomalous behavior different from that for $D_1$ and $D_2$ evolution. Also, this anomalous behavior in $D_3$ evolution differs for initial GHZ and W states. In the second case, we analyzed the influence of the initial quantum correlation on the geometric measure of quantum discord in non-Markovian regime. We observe that the effects of initial quantum correlation on TQC dynamics displays different behaviors for the two types of three qubits quantum states as expected. In the third case, we have discussed the effects of the purity $r$ on quantum correlation dynamics and compared with that of entanglement dynamics of three qubits using same conditions. This implies that quantum discord is more robust than entanglement against decoherence. Moreover, geometric measure of quantum discord have similar behavior to that for corresponding initial state in pairwise quantum discord of three qubits and that in two-qubit system.

\appendix
\section*{Appendix  } \label{app:ap}
If we have the state of three-qubits $\rho$ expressed in its Bloch representation as in Eq.(\ref{eq5}).
\begin{enumerate}
	\item  We express $\rho$ in  the orthonormal bases $\left\{X_{i_{m}}^{(m)}\right\}, i_{m}=1,2,3,4$ as the generators of $SU(2_m);\quad m=1,2,3$ labeling for qubit. $\rho_{12 \cdots N}=\sum_{i_{1} i_{2}  i_{3}} C_{i_{1} i_{2} i_{3}} X_{i_{1}}^{(1)} \otimes X_{i_{2}}^{(2)} \otimes  X_{i_{3}}^{(3)}$, where $ X_{1}^{(m)}=\frac{1}{\sqrt{2}} I_{2}$ and $X_{i_{m}}^{(m)}=\frac{1}{\sqrt{2}} \sigma_{i_{m}-1};\quad  i_{m}=2,3,4,\; \sigma_i$ being the Pauli operators \cite{has12}.
	\item We calculate the tensor $C$ of $\rho$ state as
	\begin{equation*}\label{a1}
		C_{i_1 i_2 i_3}=\\Tr\left(\rho X_{i_1}^{(1)} \otimes X_{i_2}^{(2)} \otimes X_{i_3}^{(3)}\right); i_1,i_2,i_3=1,2,3,4. ~~~~~~~~~~~~~~~~~~(A1)
	\end{equation*}
	the three-way array (tensor of order 3) with size $4\times 4\times 4$. The norm of tensor C is
	$$
	\|C\|^{2}=\sum_{i_1 i_2 i_3}^{4} C_{i_1 i_2 i_3}^{2}.
	$$
	\item We calculate $\textbf{s}^{(m)}$ a Blech vector corresponding to $m=$th qubit $m=1,2,3$,
	           \begin{eqnarray*}\label{a2}
		\textbf{s}_{i}^{(1)}&=&Tr\left(\rho \sigma_{i} \otimes I \otimes I\right),  \nonumber\\
        \textbf{s}_{j}^{(2)}&=&Tr\left(\rho I \otimes \sigma_{j} \otimes I\right),  \nonumber\\
		\textbf{s}_{k}^{(3)}&=&Tr\left(\rho I \otimes I \otimes \sigma_{k}\right),  \nonumber ~~~~~~~~~~~~~~~~~~~~~~~~~~~~~~~~~~~~~~~~~~~~~~~~~~~~(A2)
 \end{eqnarray*}
	 the correlation matrices for two-qubit as

		\begin{eqnarray*}\label{a3}
 			\mathrm{T}^{\{1,2\}}&=&[t_{i j}^{(12)}]=[Tr\left(\rho \sigma_{i} \otimes \sigma_{j} \otimes I\right)]=C_{i+1,j+1,1},\nonumber\\
			\mathrm{T}^{\{1,3\}}&=&[ t_{i k}^{(13)}]=[Tr\left(\rho \sigma_{i} \otimes I \otimes \sigma_{k}\right)]=C_{i+1,1,k+1}\nonumber\\
			\mathrm{T}^{\{2,3\}}&=&[t_{jk}^{(23)}]=[Tr\left(\rho  I \otimes \sigma_{j} \otimes \sigma_{k}\right)]=C_{1,j+1,k+1}, \nonumber
	 ~~~~~~~~~~~~~~~~~~~~~~~~~(A3)
 \end{eqnarray*}

	the three-way correlation array for $\rho$ state as
	\begin{equation*}\label{a4}
		\mathcal{T}=[t_{i j k}]=[Tr\left(\rho \sigma_{i} \otimes \sigma_{j} \otimes  \sigma_{k}\right)]=C_{i+1,j+1,k+1},\; i,j,k=1,2,3 ~~~~~~~~~~~~~~ (A4)
	\end{equation*}
	 $\mathbf{T}^{(3)}$ is $3 \times 3$ matrix, defined elementwise as
	\begin{equation*}\label{a5}
		\mathbf{T}^{(3)}=[\tau_{i\ell}^{(3)}]=[\sum_{jk} t_{ijk} t_{\ell jk}]. ~~~~~~~~~~~~~~~~~~~~~~~~ (A5)
	\end{equation*}
	\item We calculate $3 \times 3$ real symmetric matrix $G^{(1)}$ using eqs. A2, A3 and A5 as
	\begin{equation*}\label{a6}
		G^{(1)}=\textbf{s}^{(1)}(\textbf{s}^{(1)})^t+\ (\mathrm{T}^{\{1,2\}})^t \mathrm{T}^{\{1,2\}}+(\mathrm{T}^{\{1,3\}})^t \mathrm{T}^{\{1,3\}}+ \mathbf{T}^{(3)},   ~~~~~~~~~~~(A6)
	\end{equation*}

	and we find the largest eigenvalue of $G^{(1)}$, $\eta_{max}^{(1)}$ and corresponding eigenvector of it $V_{max}^{(1)}$.
	\item We calculate $D_{1}$, quantum discord (corresponding to the Von Neumann measurement on the first qubit), for a three-qubits quantum state using eqs. A2, A3 and A4 as
	\begin{equation*}
		D_{1}(\rho)  = \frac{1}{2^3} \left[ \|\textbf{s}^{(1)}\|^2+  \|\mathrm{T}^{\{1,2\}}\|^2+\|\mathrm{T}^{\{1,3\}}\|^2+\|\mathcal{T}\|^2-\eta_{max}^{(1)}\right].
	\end{equation*}
	\item We put $\hat{e}=V_{max}^{(1)}$ to calculate $\textbf{a}_{1}=\frac{1}{\sqrt{2}}(1, \hat{e})$ and $\textbf{a}_{2}=\frac{1}{\sqrt{2}}(1,-\hat{e})$.
	\item We calculate the $b_{1}$ and $b_{2}$ matrices Eqs. (13,14) in Ref. \cite{has12} of the state $\widetilde{\Pi}^{(1)}(\rho)$ using Eq. A1, $\textbf{a}_1$ and $\textbf{a}_2$ as
	$$
	b_{1 j k}=\sum_{i} c_{i j k} a_{i i}\quad,\quad b_{2 j k}=\sum_{i} c_{i j k} a_{2 i}
	$$
		\item We calculate the tensor $C^{\prime}$ of $\widetilde{\Pi}^{(1)}(\rho) =\widetilde{\rho}$  as
	$$
	C_{i j k}^{\prime}=a_{1i} b_{1 j k}+ a_{2 i} b_{2 j k}
	$$
	\textbf{proof:}

\begin{eqnarray*}\label{a7}
		C_{i j k}^{\prime}&=&Tr\left(\widetilde{\Pi}^{(1)}(\rho) X_{i} \otimes X_{j} \otimes  X_{k}\right)\nonumber\\
			&=&Tr\left(\sum_{\ell=1}^{2} p_{\ell}|\ell\rangle\langle \ell| \otimes \rho_{23/\ell}\left(X_{i} \otimes X_{j} \otimes  X_{k}\right) \right)\nonumber\\
			&=&\sum_{\ell=1}^{2}Tr\left(|\ell\rangle\langle \ell| X_{i}\right)Tr\left(p_{\ell}\rho_{23/\ell} X_{j} \otimes  X_{k}\right)\nonumber\\
			&=&\sum_{\ell=1}^{2}\left\langle \ell\left|X_{i}\right| \ell\right\rangle Tr\left(p_{\ell}\rho_{23/\ell} X_{j} \otimes  X_{k}\right)\nonumber\\
			C_{i j k}^{\prime}&=& \sum_{\ell=1}^{2} a_{\ell i} b_{\ell j k}=a_{1i} b_{1 j k}+ a_{2 i} b_{2 j k} \nonumber
		  ~~~~~~~~~~~~~~~~~~(A7)
\end{eqnarray*}
	the norm of $C_{i j k}^{\prime}$ is $\|C_{i j k}^{\prime}\|^{2}=\sum_{ijk} C_{i j k}^{\prime2}$.
	\item We repeat the step (iii) for the  state $\widetilde{\Pi}^{(1)}(\rho)=\widetilde{\rho}$, to calculate the coherent vectors $\textbf{s}^{\prime^{(m)}}$, the correlation matrix for two-qubit, three-way correlation array and $\mathbf{T}^{(3)}$ as

$$s_{j}^{\prime(1)}=Tr\left(\widetilde{\Pi}^{(1)}(\rho) \sigma_{j} \otimes I \otimes I\right)= 2^{\frac{3}{2}} C_{j+1,1,1}^{\prime}$$
$$s_{j}^{\prime(2)}=Tr\left(\widetilde{\Pi}^{(1)}(\rho) I \otimes \sigma_{j} \otimes I\right)= 2^{\frac{3}{2}} C_{1,j+1,1}^{\prime}$$
\begin{equation*}
  s_{j}^{\prime(3)}=Tr\left(\widetilde{\Pi}^{(1)}(\rho) I \otimes I \otimes \sigma_{j}\right)= 2^{\frac{3}{2}} C_{1,1j+1}^{\prime} ~~~~~~~~~~~~~~~~(A8)
\end{equation*}

and
\begin{eqnarray*}\label{a9}
			t_{i j}^{\prime(12)}&=&Tr\left(\widetilde{\Pi}^{(1)}(\rho) \sigma_{i} \otimes \sigma_{j} \otimes  I\right) =2^{\frac{3}{2}} C_{i+1,j+1,1}^{\prime}\nonumber\\
			t_{i k}^{\prime(13)}&=&Tr\left(\widetilde{\Pi}^{(1)}(\rho) \sigma_{i} \otimes I \otimes  \sigma_{k}\right) =2^{\frac{3}{2}} C_{i+1,1,k+1}^{\prime}\nonumber\\
			t_{j k}^{\prime(23)}&=&Tr\left(\widetilde{\Pi}^{(1)}(\rho) I \otimes \sigma_{j} \otimes  \sigma_{k}\right) =2^{\frac{3}{2}} C_{1,j+1,k+1}^{\prime} \nonumber
		 ~~~~~~~~~   (A9)
  \end{eqnarray*}
and
	\begin{equation*}\label{a10}
		t_{i j k}^{\prime}=\\Tr\left(\widetilde{\Pi}^{(1)}(\rho) \sigma_{i} \otimes \sigma_{j} \otimes  \sigma_{k}\right)=2^{\frac{3}{2}} C_{i+1,j+1,k+1}^{\prime} ~~~~~~~~ (A10)
	\end{equation*}
	where $ \|\mathcal{T}^{\prime}\|^2=\sum_{ijk} t_{i j k}^{\prime2},$
$$ \widetilde{\mathbf{T}}_{i\ell}^{ (3)}=\sum_{jk} t_{j i k}^{\prime} t_{j \ell k}^{\prime}. ~~~~~~~~~~~~~~~~~~~~~~~~~~~~~~~~~~~~~~~~~~~~~~~~~~~~~ (A11)
	$$
	\item We repeat steps (iv) and (v) using Eqs. A8, A9, A10 to calculate $G^{\prime(2)}$ real symmetric
	matrix for density matrix $\widetilde{\Pi}^{(1)}(\rho)$ as
	\begin{equation*}\label{a11}
		G^{\prime(2)}=S^{\prime(2)}(S^{\prime(2)})^t+\ (\mathrm{T}^{\prime\{1,2\}})^t \mathrm{T}^{\prime\{1,2\}}+(\mathrm{T}^{\prime\{2,3\}})^t \mathrm{T}^{\prime\{2,3\}}+ \mathbf{T}^{\prime(3)} , ~~~~~(A11)
	\end{equation*}
	 we find the largest eigenvalue $\eta_{\max }^{(2)}$ of $G^{\prime(2)}$ and corresponding eigenvector $V_{max }^{(2)}.$
We calculate $D_2$ as
$$D_{2}(\widetilde{\Pi}^{(1)}(\rho))  = \frac{1}{2^3} \left[ \|\textbf{s}^{\prime(2)}\|^2+  \|\mathrm{T}^{\prime\{1,2\}}\|^2+\|\mathrm{T}^{\prime\{2,3\}}\|^2+\|\mathcal{T}^{\prime}\|^2-\eta_{max}^{(2)}\right].$$
	\item We put $\hat{e}_{2}=V_{max}^{(2)}$ to calculate
	\begin{equation*}
		\tilde{\textbf{a}}_{1}=\frac{1}{\sqrt{2}}\left(1, \hat{e}_{2}\right) \quad , \quad \tilde{\textbf{a}}_{2}=\frac{1}{\sqrt{2}}\left(1,-\hat{e}_{2}\right)
	\end{equation*}
	\item We repeat the step (vii) to calculate $\widetilde{b}_{1}$ and $\widetilde{b}_{2}$ matrix of the state $\widetilde{\Pi}^{(2)}\left(\widetilde{\Pi}^{(1)}\left(\rho\right)\right)=\widetilde{\widetilde{\rho}}$.
$$
	\widetilde{b}_{1 i k}=\sum_{j} c_{i j k} \tilde{a}_{1j}\quad,\quad b_{2 i k}=\sum_{j} c_{i j k} \tilde{a}_{2j}
	$$
	
	\item We repeat the step (viii) to calculate
$$C_{i j k}^{\prime\prime}=\widetilde{a}_{1j} \widetilde{b}_{1 i k}+ \widetilde{a}_{2 j} \widetilde{b}_{2 i k}$$.

	\item We repeat the step (ix) to calculate the coherent vectors $\textbf{s}^{\prime\prime^{(m)}}$, the correlation matrix for two-qubit $t^{\prime\prime(12)}, \; t^{\prime\prime(23)}, \; t^{\prime\prime(13)}$, three-way correlation array $t^{\prime\prime}$ and $\mathbf{T}^{\prime\prime(3)}$ for state $\widetilde{\Pi}^{(2)}\left(\widetilde{\Pi}^{(1)}\left(\rho\right)\right)$.

	\item We repeat step (x) to calculate $G^{\prime\prime(3)}$ for state $\widetilde{\Pi}^{(2)}\left(\widetilde{\Pi}^{(1)}\left(\rho\right)\right)=\widetilde{\widetilde{\rho}}$ and find $\eta_{\max }^{(3)}$ and using it to calculate  $D_{3}\left(\widetilde{\Pi}^{(2)}\left(\widetilde{\Pi}^{(1)}\left(\rho\right)\right)\right).$

	\item We calculate TQC, $Q(\rho)$ as in Eq. (\ref{eq11}).
\end{enumerate}

\end{document}